\documentclass[a4paper,preprint,eqsecnum,showpacs,prb]{revtex4}
\usepackage[T1]{fontenc}
\usepackage[latin1]{inputenc}
\usepackage{graphicx}
\usepackage{amssymb}


\begin{document}

\title{Second harmonic generation in a polar ferrimagnet GaFeO$_3$}

\author{Jun-ichi Igarashi$^1$ and Tatsuya Nagao$^2$}

\affiliation{$^1$Faculty of Science, Ibaraki University, Mito,
Ibaraki 310-8512, Japan \\
$^2$Faculty of Engineering, Gunma University, Kiryu, 
Gunma 376-8515, Japan}

\begin{abstract}

We have studied second harmonic generation (SHG) in a polar ferrimagnet 
GaFeO$_3$, employing a FeO$_6$ cluster model in which the Fe atom is slightly
shifted from the center of the octahedron.
The electric-dipole transition could take place between the $3d$
states through the effective hybridization of the $4p$ states with the
$3d$ states, due to the breaking of the space-inversion symmetry. 
In the third-order perturbation with $H_{\rm int}
= -\frac{1}{c}{\bf j}\cdot{\bf A}$,
we calculate the probability per unit time, $I_{\eta aa}$,
for the process that two photons are absorbed with polarization parallel to
the $a$ axis and one photon is emitted with polarization parallel to
the $\eta$ ($=a,b,c$) axis. 
The calculated SHG intensities consist of several peaks 
as a function of two-photon energy in agreement with the experiments.
It is found that the corresponding amplitude $S_{aaa}$ at each Fe site changes
its sign while $S_{baa}$ remains the same with the reversal of
the direction of the local magnetic moment.
This implies that $I_{aaa}$ would disappear while $I_{baa}$ would survive
in the paramagnetic phase in accordance with the experiment.

\end{abstract}

\pacs{78.20.Ls, 78.20.Bh, 78.40.-q}

\maketitle

\section{Introduction}
When the space-inversion symmetry and the time-reversal symmetry
are simultaneously broken, novel magneto-optical effects are expected
to come out in the optical spectroscopy.
Such effects are known as the Kerr effect, the Faraday effect, 
the reciprocal dichroism, the magneto-chiral dichroism, and so on.
\cite{Landau,Markelov1977,Rikken1997,Rikken2002,Krichevtsov1996}

Since GaFeO$_3$ exhibits simultaneously spontaneous electric polarization 
and magnetization at low temperatures, this compound is quite suitable
to investigate such magneto-electric effects.
Remeika was the first who synthesized the compound for decades ago.
\cite{Remeika1960} 
Rado observed the large magneto-electric effect.\cite{Rado1964}
Recently, untwinned large single crystals have been prepared.\cite{Arima2004} 
The optical absorption measurement has been carried out 
with changing the direction of magnetization,\cite{Jung2004}
and the magneto-electric effects on the absorption coefficient have been
observed.
That is, the absorption spectra depend on the direction of 
the magnetization in the region of photon energy $1.0-2.5$ eV.
In our previous paper,\cite{Igarashi2009}
we have analyzed the spectra through the microscopic calculation.
The calculated spectral shape has agreed well with the experimental curve.
\cite{Jung2004}

In the field of the nonlinear optics, it is known that the breaking of
the spatial inversion symmetry in polar or chiral materials gives rise to
second harmonic generation (SHG).\cite{Bloembergen,Shen,Fiebig2005}
When the time-reversal symmetry is simultaneously broken
in polar magnetic materials, the SHG intensities with some specific
polarization are activated by the presence of magnetization.
\cite{Fiebig1994,Froehlich1998,Fiebig2002,Pisarev2004,Meier2009}
In this context, the SHG spectra have been observed and analyzed in plenty of
systems such as Cr$_2$O$_{3}$ and multiferroic
materials,\cite{Fiebig1994,Fiebig1997,Muto1998,Tanabe1998,
Nogami2008,Lottermoser2009,Zimmermann2009}
and magnetic field induced SHG has been investigated
in several semiconductors.\cite{Ogawa2004.2,Pavlov2005,Sanger2006}
Recently, the SHG experiments have been carried out in GaFeO$_3$.
\cite{Ogawa2004,Eguchi2005,Kalashnikova2005,Matsubara2009}
Ogawa \textit{et al}. have measured the magnetization-induced SHG spectra
of GaFeO$_{3}$ 
in the region of the two-photon energy $2.5-4.5$ eV.\cite{Ogawa2004}
The purpose of this paper is to elucidate the origin of the SHG spectra
through the microscopic calculation.

It is known that the crystal of GaFeO$_3$ belongs to the space group 
$Pc2_{1}n$ and has an orthorhombic unit cell.\cite{Wood1960}
Each Fe atom is octahedrally surrounded by O atoms,
and is slightly displaced from the center of the octahedron
along the $b$ axis.
There are ``Fe1" and ``Fe2" sites for Fe atoms, where the displacement
is $0.26 \textrm{\AA}$ at Fe1 sites and $-0.11 \textrm{\AA}$ at Fe2 sites.
\cite{Arima2004} 
Thereby the spontaneous electric polarization is generated along the $b$ axis. 
The local magnetic moments at Fe1 and Fe2 sites are known to align 
antiferromagnetically along the $\pm c$ axis. 
Note that the actual compound deviates slightly
from a perfect antiferromagnet and behaves as a ferrimagnet.\cite{Frankel1965} 
The origin for this deviation is not fully understood, but is inferred that 
the Fe occupation at Fe1 and Fe2 sites are slightly different from each other.
\cite{Arima2004}
We assume the system as a perfect antiferromagnet in the following analysis.
As shown later, we could obtain the SHG intensities in the antiferromagnetic 
phase, since the system is breaking the space-inversion symmetry. 
This situation is different from those of EuTe and EuSe,\cite{Kaminski2009}
where no SHG intensities exist in the antiferromagnetic phase 
because Eu atoms reside on centrosymmetric sites.
We expect that the slight deviation from the perfect antiferromagnet would
cause merely minor quantitative change in the SHG intensities.

We introduce a FeO$_6$ cluster model in which the Fe atom is slightly displaced
from the center of the octahedron.\cite{Igarashi2009}
We neglect the slight distortion of octahedron. 
We take account of the Coulomb interaction between the $3d$ states,
the spin-orbit interaction on the $3d$ states, and the hybridization of
the oxygen $2p$ states with the Fe $3d$ and $4p$ states.
The same cluster model has been successfully applied to investigating the
magneto-electric spectra in the optical absorption\cite{Igarashi2009}
and the directional dichroic spectra in the $K$-edge x-ray absorption
\cite{Igarashi2010} in GaFeO$_3$.
We evaluate the matrix elements of the electric dipole
($E1$) transition between the $3d$ and the $4p$ states using the \emph{atomic}
Hamiltonian from the conventional form 
$H_{\rm int}=-\frac{1}{c}{\bf j}\cdot {\bf A}$ where
${\bf j}$ is the current operator and ${\bf A}$ is the vector potential.
As shown in Sec.~III, the matrix elements thus evaluated are found larger 
than those evaluated from another conventional
form $H_{\rm int}=-{\bf P}\cdot{\bf E}$ where
${\bf P}$ is the electric dipole operator and ${\bf E}$ is the electric field. 
The situation may become different without using the atomic Hamiltonian 
in the band structure calculation.
The $E1$ transition between the $3d$ states could eventually take place
through the effective $4p$-$3d$ hybridization due to the breaking of the 
space-inversion symmetry. Thereby the effective $E1$ transition matrix 
elements become larger than those of the magnetic dipole ($M1$) transition
in the $3d^5$ configuration. In the present cluster model analysis,
we regard $H_{\rm int}=-\frac{1}{c}{\bf j}\cdot {\bf A}$ more fundamental.

Bearing the above situation in mind, we avoid to use the form 
$H_{\rm int}=-{\bf P}\cdot{\bf E}$ in the cluster model analysis.
Then, using perturbation theory to third-order with 
$H_{\rm int}=-\frac{1}{c}{\bf j}\cdot{\bf A}$,
we formulate the SHG intensity
from the probability per unit time of the process that the incident 
two photons with the frequency $\omega$ are absorbed and 
one photon with the frequency $2 \omega$ is emitted.
Although this approach seems different  
from the conventional analysis using the nonlinear susceptibility, 
the expression for the probability amplitude is quite close to that obtained
from the nonlinear susceptibility. The difference is that the $E1$ transition 
matrix elements in the formula are not evaluated from 
$H_{\rm int}=-{\bf P}\cdot{\bf E}$.
On the basis of this formula on the cluster model, we calculate the 
SHG intensities as a function of two-photon energy on various polarization 
conditions in the so-called phase-matching condition.
The $M1$ transition is found to give a minor contribution to the SHG intensity.
The calculated SHG spectra show multi-peak structure in the 
$1.0$-$4.5$ eV range in agreement with the available experimental data. 
Another finding is that, when the polarizations of both the incident 
and emitted photons are all parallel to the $a$ axis, the amplitude 
(complex number) at each Fe site change its sign with the reversal of 
the direction of the local magnetization.
This indicates that the SHG intensity, which will be
denoted as $I_{aaa}$, would disappear by passing through from the 
antiferromagnetic phase to the paramagnetic phase. 
Therefore, this spectrum may be called as magneto-electric one
because it is activated by the magnetization.
On the other hand, we confirm that, when the polarization of incident 
photons is parallel to the $a$ axis while that of the emitted photon is 
parallel to the $b$ axis, the corresponding amplitude at each Fe site keeps 
the same value with the reversal of the direction of the local magnetic moment.
This implies that the SHG intensity, which will be denoted as $I_{baa}$, 
would change little by passing through
from the antiferromagnetic phase to the paramagnetic phase. 
These characteristics of the polarization dependence 
as well as the spectral shape agree with the SHG spectra observed in the 
reflection measurement,\cite{Ogawa2004,Matsubara2009}
although our results is not for the reflection spectra.

This paper is organized as follows. In Sec. II, we introduce a cluster model
FeO$_6$. In Sec. III, we describe the optical transition operators
associated with Fe atoms.
In Sec. IV, we derive the formula of the SHG intensity, and
present the calculated spectra in comparison with the experiment.
The last section is devoted to concluding remarks.

\section{Hamiltonian for a F\lowercase{e}O$_6$ cluster}

Considering a FeO$_6$ cluster, here we briefly summarize the corresponding 
model Hamiltonian.
The details are found in our previous papers.\cite{Igarashi2009,Igarashi2010}
It may be expressed as
\begin{equation}
 H = H^{3d} + H^{2p} + H^{4p} 
 + H_{\rm hyb}^{3d-2p} + H_{\rm hyb}^{4p-2p}. \label{eq.Ham}
\end{equation}
The $H^{3d}$ represents the energy of Fe $3d$ electrons, which includes
the intra-atomic Coulomb interaction expressed in terms of the Slater 
integrals, the spin-orbit interaction, and the energy arising from 
the exchange interaction via the exchange field from neighboring Fe atoms.
The energy of the Fe $4p$ states
and that of the oxygen $2p$ states
are represented by $H^{4p}$ and $H^{2p}$, respectively.

The $H^{3d-2p}_{\rm hyb}$ and $H^{4p-2p}_{\rm hyb}$ represent
the hybridization energies of the
O $2p$ states with the Fe $3d$ and $4p$ states, respectively:
\begin{eqnarray}
 H_{\rm hyb}^{3d-2p} &=& \sum_{j\eta\sigma m}t_{m\eta}^{3d-2p}(j)
   d_{m\sigma}^{\dagger}p_{j\eta\sigma}+{\rm H.c.}, \\
 H_{\rm hyb}^{4p-2p} &=& \sum_{j\eta\sigma\eta'} 
 t_{\eta'\eta}^{4p-2p}(j)
 p'^{\dagger}_{\eta'\sigma} p_{j\eta\sigma}+{\rm H.c.},
\end{eqnarray}
where $d_{m\sigma}^{\dagger}$ and $p'^{\dagger}_{\eta\sigma}$ 
stand for creation
operators of electron with the $3d$ orbital ($m=x^2-y^2, 3z^2-r^2, yz, zx$,
and $xy$) having spin $\sigma$
and of the local $4p$ orbital ($\eta=x, y$, and $z$) having spin $\sigma$,
respectively.  
The $p_{j\eta\sigma}$ is the annihilation operator of electron with the
oxygen $2p$ orbital at neighboring site $j$.
The sum over $j$ is taken on neighboring O sites.
The hybridization parameters $t^{3d-2p}_{m\eta}(j)$ and 
$t^{4p-2p}_{\eta'\eta}(j)$ are expressed in terms of the Slater-Koster
two-center integrals.

The Fe atom is slightly displaced from the center of the octahedron;
the shift $\delta$ is $0.26 \textrm{\AA}$ at Fe1 sites 
and $-0.11 \textrm{\AA}$ at Fe2 sites along the $b$ axis. Therefore,
the hybridization parameters are modified.
We evaluate the modified Slater-Koster two-center integrals for the Fe atom 
by assuming that $(pd\sigma)_{2p,3d}$, 
$(pd\pi)_{2p,3d}\propto d^{-4}$, and $(pp\sigma)_{4p,2p}$, 
$(pp\pi)_{4p,2p}\propto d^{-2}$ for $d$ being the Fe-O distance.
\cite{Harrison2004}
With these modified values, we obtain the ligand field Hamiltonian 
on the $3d$ states, which is deviated from the cubic symmetry.
In the second-order perturbation, it may be given by 
\begin{equation}
 \tilde{H}^{3d-3d}= \sum_{mm'\sigma} \tilde{t}_{mm'}^{3d-3d} 
  d_{m\sigma}^{\dagger}d_{m'\sigma},
\end{equation}
with
\begin{equation}
 \tilde{t}^{3d-3d}_{mm'} = \sum_{j\eta} t^{3d-2p}_{m\eta}(j)
 t^{3d-2p}_{m'\eta}(j)/\Delta,
\end{equation}
where $\Delta$ denotes the charge transfer energy.
In addition to the ligand field Hamiltonian, we have
the effective hybridization between the $4p$ and $3d$ states,
due to the breaking of the space-inversion symmetry.
In the second-order perturbation, the hybridization energy may be given by
\begin{equation}
 \tilde{H}^{4p-3d}= \sum_{\eta'm\sigma} \tilde{t}_{\eta'm}^{4p-3d} 
  p'^{\dagger}_{\eta'\sigma}d_{m\sigma} + {\rm H.c.},
\end{equation}
with
\begin{equation}
 \tilde{t}^{4p-3d}_{\eta'm} = \sum_{j\eta} \frac{t^{4p-2p}_{\eta'\eta}(j)
       t^{3d-2p}_{m\eta}(j)}{E^{4p}-E^{2p}},
\label{eq.2.7}
\end{equation}
where $E^{4p}$ and $E^{2p}$ are the average of the $4p$-band energy
and the energy of the O $2p$ electron, respectively.
The denominator in Eq. (\ref{eq.2.7}) is approximately estimated
as $E^{4p}-E^{2p}\approx 17$ eV. 

In the following numerical calculation, we use the same parameter values as 
in our previous papers,\cite{Igarashi2009,Igarashi2010}
except for $\Delta=4.0$ eV, 
which is slightly larger than the previous value 3.3 eV.

\section{Interaction between electromagnetic wave and electron\label{sect.3}}

We concentrate our attention on Fe atoms. Then,
the interaction between electrons and the electromagnetic wave 
with polarization vector ${\bf e}$ and wave vector ${\bf q}$ 
is approximated as
\begin{equation}
 H_{\rm int} = -\frac{1}{c}\sum_{i}{\bf j}({\bf q},i)\cdot
 {\bf A}({\bf q},i) + {\rm H.c.},
\label{eq.jA}
\end{equation}
with
\begin{eqnarray}
 {\bf j}({\bf q},i) &=& \sum_{nn'} 
  \left[\int {\rm e}^{i{\bf q}\cdot({\bf r}-{\bf r}_i)}
  {\bf j}_{nn'}({\bf r}-{\bf r}_i){\rm d}^3({\bf r}-{\bf r}_i)\right]
    a_{n}^{\dagger}(i)a_{n'}(i) , 
\label{eq.current}\\
 {\bf A}({\bf q},i) &=& \sqrt{\frac{2\pi\hbar c^2}{V\omega_{\bf q}}}
   {\bf e}c_{\bf q}{\rm e}^{i \textbf{q}\cdot\textbf{r}_i},
\end{eqnarray}
where ${\bf r}_i$ is the position vector of the Fe atom at site $i$, 
and $a_n(i)$ is the annihilation operator of electron with the local 
wave function $\phi_n({\bf r}-{\bf r}_i)$. 
The ${\bf j}_{nn'}({\bf r}-{\bf r}_i)$ in Eq.~(\ref{eq.current})
is given by
\begin{eqnarray}
 {\bf j}_{nn'}({\bf r}-{\bf r}_i) &=& \frac{ie\hbar}{2m}
 \big[(\nabla \phi^{*}_n)\phi_{n'}
  - \phi_n^{*}\nabla\phi_{n'}\big] - \frac{e^2}{mc}{\bf A}\phi_n^{*}\phi_{n'}¡
  \nonumber\\
  &+& \frac{e\hbar}{mc}c\nabla\times [\phi_n^{*}{\bf S}\phi_{n'}],
\label{eq.localcurrent}
\end{eqnarray}
where $e$, $m$ and $\hbar{\bf S}$ mean the charge, the mass, 
and the spin operator of electron, respectively.
The interaction Hamiltonian can be rewritten as
\begin{equation}
 H_{\rm int} = -e\sqrt{\frac{2\pi}{V\hbar\omega_{\bf q}}}
    \sum_{i} T({\bf q},{\bf e},i)
     c_{\bf q} {\rm e}^{i \textbf{q}\cdot\textbf{r}_i} + {\rm H.c.} ,
\end{equation}
where the transition operator $T({\bf q},{\bf e},i)$ is
defined as $\hbar \textbf{e} \cdot \textbf{j}(\textbf{q},i)/e$.

First, we consider the $E1$ transition.
Putting ${\rm e}^{i{\bf q}\cdot({\bf r}-{\bf r}_j)}=1$ 
in Eq.~(\ref{eq.current}), we evaluate  the first term in 
Eq.~(\ref{eq.localcurrent}) by using the relation
\begin{equation}
 \int \phi_n^* \frac{\partial}{\partial z}\phi_{n'} 
{\rm d}^3{\bf r} = -\frac{m}{\hbar^2}(\epsilon_n-\epsilon_{n'})
  \int \phi_n^* z\phi_{n'} {\rm d}^3{\bf r},
\end{equation}
where $\epsilon_n$ and $\epsilon_{n'}$ describe the 
energy eigenvalues with eigenfunctions $\phi_n$ and 
$\phi_{n'}$, respectively. 
In the present study, $\phi_n$ and $\phi_{n'}$ are assigned to the $4p$ 
and $3d$ states.
Then, the transition operator $T^{E1}$ is summarized as 
\begin{equation}
 T^{E1}({\bf q},{\bf e},i) =
  iB^{E1} \sum_{i\eta m\sigma} N_{\eta m}^{E1}
     [p'^{\dagger}_{\eta\sigma}(i)d_{m\sigma}(i)
     -d_{m\sigma}^{\dagger}(i)p'_{\eta\sigma}(i)], 
\label{eq.e1mat}
\end{equation}
with
\begin{equation}
 B^{E1} = (\epsilon_{4p}-\epsilon_{3d})\int_{0}^{\infty}
   r^3 R_{4p}(r)R_{3d}(r){\rm d}r, 
\label{eq.dipole}
\end{equation}
where $R_{3d}(r)$, $R_{4p}(r)$ are radial wave-functions 
of $3d$, $4p$ states with energy $\epsilon_{3d}$, $\epsilon_{4p}$ 
in the Fe atom.  Within the HF approximation 
in the 1s$^2$3d$^5$4p$^{0.001}$-configuration of an Fe atom,\cite{Cowan1981} 
we estimate it as $B^{E1} \approx 7.7\times 10^{-8}$ cm$\cdot$eV.
Coefficient $N_{\eta m}^{E1}$'s are given by
$N_{x,x^2-y^2}^{E1}=1/\sqrt{5}$, 
$N_{x,3z^2-r^2}^{E1}=-1/\sqrt{15}$, 
$N_{y,xy}^{E1}=1/\sqrt{5}$, 
$N_{z,zx}^{E1}=1/\sqrt{5}$ for polarization parallel to the $x$ axis, 
$N_{x,xy}^{E1}=1/\sqrt{5}$, 
$N_{y,x^2-y^2}^{E1}=-1/\sqrt{5}$, 
$N_{y,3z^2-r^2}^{E1}=1/\sqrt{15}$, 
$N_{z,yz}^{E1}=1/\sqrt{5}$ for polarization parallel to the $y$ axis,
and
$N_{x,zx}^{E1}=1/\sqrt{5}$, 
$N_{y,yz}^{E1}=1/\sqrt{5}$, 
$N_{z,3z^2-r^2}^{E1}=2/\sqrt{15}$ for polarization parallel to the $z$ axis, 
respectively.

The matrix elements of the $E1$ transition are sometimes evaluated from 
the dipole interaction,
\begin{equation}
  \tilde{H}_{\rm int}=-\sum_{i}{\bf P}_i\cdot{\bf E}({\bf q},i),
\label{eq.dipole2}
\end{equation}
where ${\bf P}_i$ is the dipole operator for electrons of the Fe atom
at site $i$, and ${\bf E}({\bf q},i)$ is the electric field.
In Eq.~(\ref{eq.dipole2}), the matrix element of the dipole operator 
between the $4p$ and $3d$ states may be estimated as 
Eq.~(\ref{eq.dipole}) divided by $\epsilon_{4p}-\epsilon_{3d}$, 
while ${\bf E}({\bf q},i)=-\frac{1}{c}\frac{\partial{\bf A}({\bf q},i)}
{\partial t}=\frac{i\omega}{c}A({\bf q},i)$ 
for the oscillating field with the frequency $\omega$. 
Therefore, the use of Eq.~(\ref{eq.dipole2}) underestimates the $E1$-transition
matrix elements by a factor $\hbar\omega/(\epsilon_{4p}-\epsilon_{3d})$
with $\epsilon_{4p}-\epsilon_{3d} > 10$ eV and $\hbar\omega \sim$ several eV's.
We think Eq.~(\ref{eq.jA}) more fundamental from the microscopic standpoint. 

Now we evaluate the matrix elements of the $E1$-transition 
between the $3d$ states.
The matrix elements between the $3d$ states take finite values when
the $3d$ states mix with the $4p$ states.
As a first step of the evaluation, we calculate the energy eigenstates 
$|\Phi_n(d^5)\rangle$ with eigenenergy $E_n(d^5)$ in the $3d^5$-configuration,
and $|\Phi_n(d^4)\rangle$ with eigenenergy $E_n(d^4)$ in the 
$3d^4$-configuration, by diagonalizing the Hamiltonian 
$H_{3d}+\tilde{H}^{3d-3d}$. 
If the displacement is neglected, Fe atoms are under the cubic symmetry.
The displacement gives rise to an additional trigonal field,
which makes the energy levels split further. The spin-orbit interaction 
and the exchange field further modify these states. 
Note that the matrix elements of the $E1$ transition do not exist
between these states.

Next, we treat the effective hybridization $\tilde{H}^{4p-3d}$ 
within the first order perturbation.
The modified wave function $|\Psi_n(i)\rangle$ may be written as
\begin{eqnarray}
 |\Psi_n(i)\rangle &=& |\Phi_n(d^5)\rangle \nonumber \\
               &+& \sum_{m{\bf k}\eta\sigma}
  |\Phi_{m}(d^4),{\bf k}\eta\sigma\rangle 
  \frac{\langle\Phi_{m}(d^4),{\bf k}\eta\sigma| 
  \tilde{H}^{4p-3d}|\Phi_{n}(d^5)\rangle}
       {E_{n}(d^5)-[E_{m}(d^4)+\epsilon_{4p}({\bf k})]}, 
\label{eq.final.opt}
\end{eqnarray}
where $|\Phi_m(d^4),{\bf k}\eta\sigma\rangle$ represents the state of 
four electrons in the 3d states and one electron in
the $4p$ states specified by $\eta$($=x,y$, and $z$), spin $\sigma$,
and momentum ${\bf k}$. 
The sum over ${\bf k}$ may be replaced by the integral with the $4p$ DOS,
which is explicitly given in Ref.~\onlinecite{Igarashi2009}.
>From Eq.~(\ref{eq.final.opt}), the matrix elements of the $E1$ transition 
between the states in the $3d^5$-configuration are given by
\begin{eqnarray}
 [T^{E1}({\bf q},{\bf e},i)]_{n',n}
 &\equiv& \langle\Psi_{n'}(i)|T^{E1}({\bf q},{\bf e},i)|\Psi_{n}(i)\rangle 
  \nonumber \\
  &=& \sum_{m{\bf k}\eta\sigma}  
  \frac{\langle\Phi_{n'}(d^5)|T^{E1}({\bf q},{\bf e},i)
      |\Phi_m(d^4),{\bf k}\eta\sigma\rangle
     \langle\Phi_m(d^4),{\bf k}\eta\sigma|\tilde{H}^{4p-3d}|\Phi_n(d^5)\rangle}
       {E_n(d^5)-E_m(d^4)-\epsilon_{4p}({\bf k})} 
 \nonumber\\
  &+& \sum_{m{\bf k}\eta\sigma} 
  \frac{\langle\Phi_{n'}(d^5)|\tilde{H}^{4p-3d} 
      |\Phi_m(d^4),{\bf k}\eta\sigma\rangle 
      \langle\Phi_m(d^4),{\bf k}\eta\sigma|T^{E1}({\bf q},{\bf e},i)
      |\Phi_n(d^5)\rangle}
       {E_{n'}(d^5)-E_m(d^4)-\epsilon_{4p}({\bf k})}. \nonumber \\
\label{eq.ME1}
\end{eqnarray}

Finally, we close this section by evaluating the matrix elements of the
$M1$ transition.
In doing so, we approximate the third term in Eq.~(\ref{eq.localcurrent}) as
\begin{eqnarray}
\int {\rm e}^{i{\bf q}\cdot({\bf r}-{\bf r}_i)} 
 \nabla\times(\phi_n^{*}{\bf S}\phi_{n'}) {\rm d}^3{\bf r} &=&
 -i{\bf q}\times \int \phi_n^{*}{\bf S}\phi_{n'} 
 {\rm e}^{i{\bf q}\cdot({\bf r}-{\bf r}_i)} {\rm d}^3{\bf r} 
\nonumber \\
&\approx& 
 -i{\bf q}\times \int \phi_n^{*}{\bf S}\phi_{n'} {\rm d}^3{\bf r}.
\label{eq.spin}
\end{eqnarray}
In addition to this term, we have, from the first term of 
Eq.~(\ref{eq.localcurrent}), the similar term to Eq.~(\ref{eq.spin}),
in which ${\bf S}$ is replaced by ${\bf L}/2$ (${\bf L}$ is the orbital 
angular momentum). See Ref.~\onlinecite{Igarashi2009} for the derivation. 
Hence, the corresponding transition operator may be expressed as
\begin{equation}
 T^{M1}({\bf q},{\bf e},i) = 
   i|\textbf{q}|B^{M1}\sum_{imm'\sigma\sigma'} N^{M1}_{m\sigma,m'\sigma'}
   d^{\dagger}_{m\sigma}(i)d_{m'\sigma'}(i),
\label{eq.m1mat}
\end{equation}
where $B^{M1}=\hbar^2/2m=3.8\times 10^{-16}{\rm cm}^2\cdot{\rm eV}$.
When the photon propagates along the $z$ axis,
$N^{M1}_{m\sigma,m'\sigma'}=\langle m\sigma|-(L_y+2S_y)|m'\sigma'\rangle$
for polarization parallel to the $x$ axis, and
$N^{M1}_{m\sigma,m'\sigma'}=\langle m\sigma|L_x+2S_x|m'\sigma'\rangle$
for polarization parallel to the $y$ axis.
The matrix elements between the $3d$ states in the $d^5$-configuration
are given by
\begin{eqnarray}
 [T^{M1}({\bf q},{\bf e},i)]_{n',n}
 &\equiv& \langle\Psi_{n'}(i)|T^{M1}({\bf q},{\bf e},i)|\Psi_{n}(i)\rangle 
 \nonumber \\
&\approx& \langle\Phi_{n'}(d^5)|T^{M1}({\bf q},{\bf e},i)|\Phi_{n}(d^5)\rangle .
\end{eqnarray}
These values are found much smaller than the matrix elements of the $E1$
transition given by Eq.~(\ref{eq.ME1}).

\section{Second Harmonic Generation}

\begin{figure}[h]
\begin{center}\includegraphics[%
  scale=0.5]{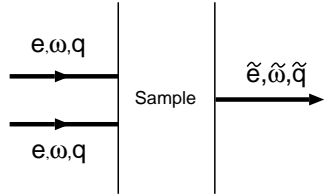}
\caption{
Schematic description of second harmonic generation.
Two photons with ${\bf e}$, $\omega$, ${\bf q}$ are absorbed,
and one photon with $\tilde{\bf e}$, $\tilde{\omega}$, $\tilde{\bf q}$ 
is emitted.
\label{fig.geometry}}
\end{center}
\end{figure}

We consider the process that two photons with the 
frequency $\omega$, wave vector 
${\bf q}$, and polarization ${\bf e}$ are absorbed and one photon with 
$\tilde{\omega}$, $\tilde{\bf q}$, and $\tilde{\bf e}$
is emitted with $\tilde{\omega}=2\omega$, as illustrated 
in Fig.~\ref{fig.geometry}. 
In the third-order perturbation with 
$H_{\rm int}=-\frac{1}{c}{\bf j}\cdot{\bf A}$, the probability per unit time
for the process may be expressed as
\begin{equation}
 I({\bf e},\omega,{\bf q};\tilde{\bf e},\tilde{\omega},\tilde{\bf q})
  \propto \left| \sum_{i}
  S({\bf e},\omega,{\bf q};\tilde{\bf e},\tilde{\omega},\tilde{\bf q};i)
  {\rm e}^{i(2{\bf q}-\tilde{\bf q})\cdot {\bf r}_i} \right|^2
  \tilde{\omega}^2 \delta(\tilde{\omega}-2\omega),
\label{eq.shg}
\end{equation}
where the amplitude $S$ is given by
\begin{eqnarray}
 &&S({\bf e},\omega,{\bf q};\tilde{\bf e},\tilde{\omega},\tilde{\bf q};i) 
 \nonumber\\
 &\propto& \frac{1}{\omega\sqrt{\tilde{\omega}}} \sum_{n',n} 
  \Biggl\{
      \frac{[T^{*}(\tilde{\bf q},\tilde{\bf e},i)]_{g,n'}
            [T({\bf q},{\bf e},i)]_{n',n}
            [T({\bf q},{\bf e},i)]_{n,g}}
           {(\epsilon_{n'}-\epsilon_{g}-2\hbar\omega-i\Gamma)
            (\epsilon_{n} -\epsilon_{g}- \hbar\omega-i\Gamma)}
     \nonumber\\
 &+&\frac{[T({\bf q},{\bf e},i)]_{g,n'}
          [T^{*}(\tilde{\bf q},\tilde{\bf e},i)]_{n',n}
          [T({\bf q},{\bf e},i)]_{n,g}}
           {(\epsilon_{n'}-\epsilon_{g}+\hbar\tilde{\omega}
	     -\hbar\omega-i\Gamma)
            (\epsilon_{n} -\epsilon_{g}-\hbar\omega-i\Gamma)}
     \nonumber\\
 &+&\frac{[T({\bf q},{\bf e},i)]_{g,n'}
          [T({\bf q},{\bf e},i)]_{n',n}
          [T^{*}(\tilde{\bf q},\tilde{\bf e},i)]_{n,g}}
           {(\epsilon_{n'}-\epsilon_{g}+\hbar\tilde{\omega}
	     -\hbar\omega-i\Gamma)
            (\epsilon_{n} -\epsilon_{g} + \hbar\tilde{\omega}
	     -i\Gamma)} \Biggr\}.
\label{eq.smatrix}
\end{eqnarray}
The $\Gamma$ represents the life-time broadening width by other random 
perturbation on the material system. 
The $\epsilon_g$ is the energy of the ground state 
in the $3d^5$-configuration, and $\epsilon_n$ is the abbreviation of 
$E_n(d^5)$.
Note that Eq. (\ref{eq.smatrix}) resembles the conventional expression of 
nonlinear susceptibility, which is based on the interaction
$\tilde{H}_{\rm int}=-\sum_{i}{\bf P}_i\cdot{\bf E}({\bf q},i)$.
The sum over $i$ in Eq.~(\ref{eq.shg}) is made on all Fe sites, and
the so-called phase-matching condition $\tilde{\bf q}=2{\bf q}$ has to be 
satisfied in order to get finite intensities. 
Note also that, if the wave interaction length $\ell$ is finite, 
the momentum conservation would be relaxed within $\Delta\tilde{\bf q}\sim
\frac{1}{\ell}$. In reflection, the surface layer with $1/q$ thickness 
could contribute significantly to generating reflection wave without phase 
matching-condition.\cite{Shen} Since the matrix elements of the $E1$
transition is much larger than those of the $M1$ transition as discussed
in Sec. III, all the transition matrix-elements in Eq.~(\ref{eq.smatrix})
could be replaced by those of the $E1$ transition.

\begin{figure}[h]
\begin{center}\includegraphics[%
  scale=0.8]{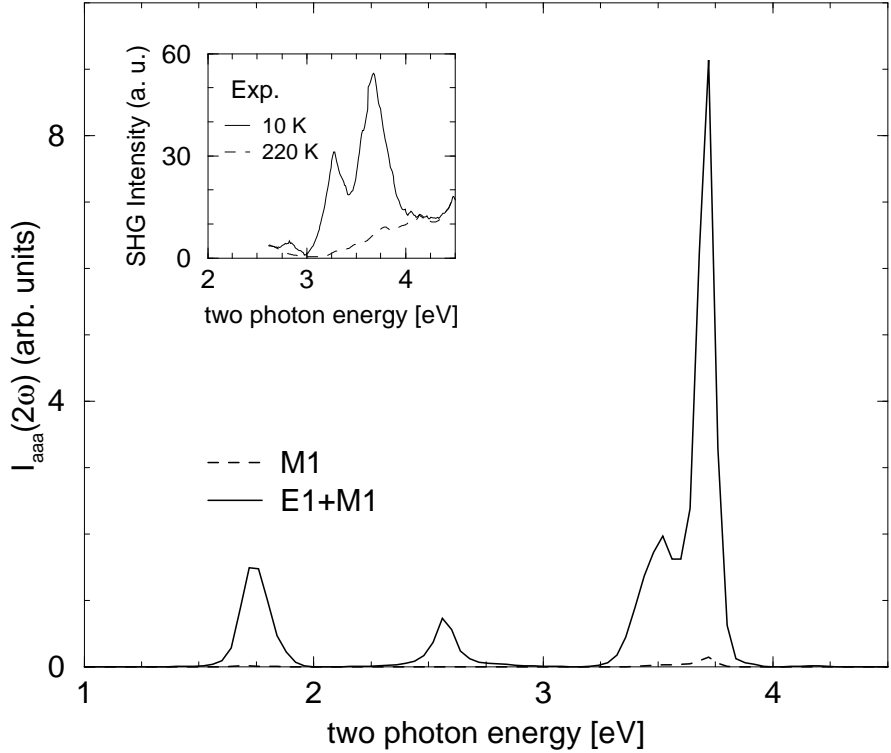}
\caption{
The second harmonic generation (SHG) intensity $I_{aaa}(2\omega)$
as a function of two-photon energy 
$2\omega$ with ${\bf e}$ and $\tilde{\bf e}$ parallel to
the $a$ axis.
Inset: experimental SHG intensity in the reflection geometry;
the $s$-polarized light is radiated with incident angle $5$ degree
on the surface of $ac$ plane, and the $s$-polarized reflected SHG light 
is observed (Ref.~\onlinecite{Ogawa2004}).
\label{fig.shg1}}
\end{center}
\end{figure}

We consider the situation that the polarization of incident photons is
parallel to the $a$ axis. When the polarization of the emitted photon 
is parallel
to the $c$ axis, the SHG intensity, which will be denoted as 
$I_{caa}(2\omega)$, is found to vanish. On the other hand, 
when the polarization of the emitted photon is parallel to the $a$ axis, 
non-zero SHG intensity $I_{aaa}(2\omega)$ is obtained
as shown in Fig.~\ref{fig.shg1}.
The broken line represents the result with replacing 
$T^{*}(\tilde{\bf q},\tilde{\bf e},i)$ by 
$T^{M1*}(\tilde{\bf q},\tilde{\bf e},i)$ in Eq.~(\ref{eq.smatrix}),
demonstrating that the contribution of the $M1$ transition is much 
smaller than that of the $E1$ transition as estimated in Sec. \ref{sect.3}.
The spectral shape is composed of multi-peak structure:
a small peak around $2\omega=1.75$ eV,
a small peak around $2.5$ eV, and a two-peak 
structure around $\hbar\omega=3-4$ eV. 
The two-peak structure reasonably captures a whole aspect of 
the experimental SHG intensity shown in the inset of Fig. \ref{fig.shg1},
although the experimental spectrum is obtained on the 
"$S_{\rm in}S_{\rm out}$" configuration
in the reflection geometry.\cite{Ogawa2004} 

\begin{figure}[h]
\begin{center}\includegraphics[%
  scale=0.8]{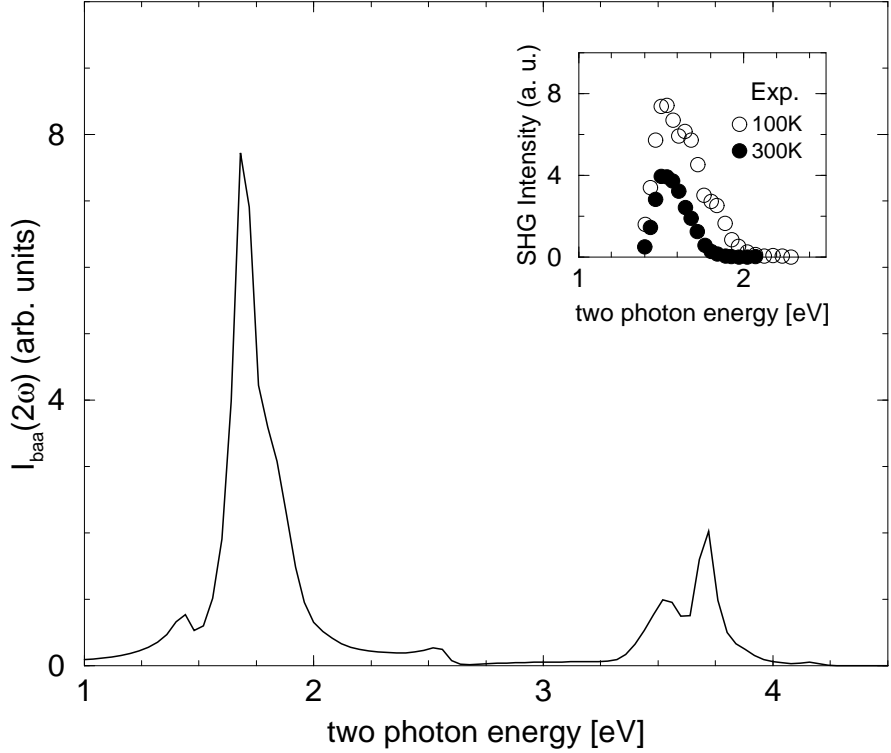}
\caption{
The SHG intensity $I_{baa}(2\omega)$
as a function of two-photon energy
$2\omega$ with ${\bf e}$ and $\tilde{\bf e}$ parallel to the $a$ and
$b$ axes, respectively.
Inset: experimental SHG intensity (Ref.~\onlinecite{Eguchi2005}).
\label{fig.shg2}}
\end{center}
\end{figure}

For the polarization of the emitted photon parallel to the $b$ axis,
we obtain the spectra $I_{baa}(2\omega)$ as shown in Fig.~\ref{fig.shg2}.
In comparison with the spectral shape of $I_{aaa}(2\omega)$, the peak around
$2\omega=1.75$ eV becomes larger, but the other peaks remain similar to 
those of $I_{aaa}(2\omega)$. As shown in the inset of Fig. \ref{fig.shg2}, 
a large peak is observed
around $2\omega=1.5$ eV in the SHG experiment,\cite{Eguchi2005}
which is reproduced by the present theory.
Our calculation also implies that another peak in the $3.3-4.0$ eV range is
anticipated if corresponding experiment will be available.
Note that the SHG intensities have been observed around $2.5-4$ eV 
on the "$S_{in}P_{\rm out}$" configuration in the reflection experiment.
\cite{Ogawa2004,Matsubara2009}
The obtained intensity is about $1/4$ of $I_{aaa}(2\omega)$ in this energy
region, corresponding well to the ratio of intensity on the 
"$S_{in}P_{\rm out}$" configuration to that on the "$S_{in}S_{\rm out}$" 
configuration in the reflection experiment.

Now we examine what happens to the SHG intensity when
the system undergoes a phase transition into the paramagnetic phase.
The probability amplitude $S_{aaa}(2\omega;i)$ is found to change its sign,
when the direction of the local magnetic moment is reversed.
This indicates that $\sum_{i}S_{aaa}(2\omega;i)$ would be canceled out
when the local magnetic moment is randomly oriented,
and that $I_{aaa}(2\omega)$ would disappear in the paramagnetic phase,
in agreement with the experiment.
For this reason, $I_{aaa}(2\omega)$ may be called as the
\textit{magnetization-induced} SHG intensity,
which corresponds well to the spectra around $2\hbar\omega\sim 3-4$ eV 
for the "$S_{\rm in}S_{\rm out}$" configuration in the reflection experiments.
\cite{Ogawa2004,Matsubara2009}
On the other hand, the amplitude $S_{baa}(2\omega;i)$ 
is found to remain the same when the direction of the local magnetic moment 
is reversed.
This leads to that $\sum_{i}S_{baa}(2\omega;i)$ would \emph{not} be
canceled out when the local magnetic moment is randomly oriented,
and that $I_{baa}(2\omega)$ remains finite in the paramagnetic phase,
which corresponds well to the spectra around $2\hbar\omega\sim 3-4$ eV 
for the ``$S_{\rm in}P_{\rm out}$" configuration in the reflection experiments.
\cite{Ogawa2004,Matsubara2009}

\begin{figure}[h]
\begin{center}\includegraphics[%
  scale=0.8]{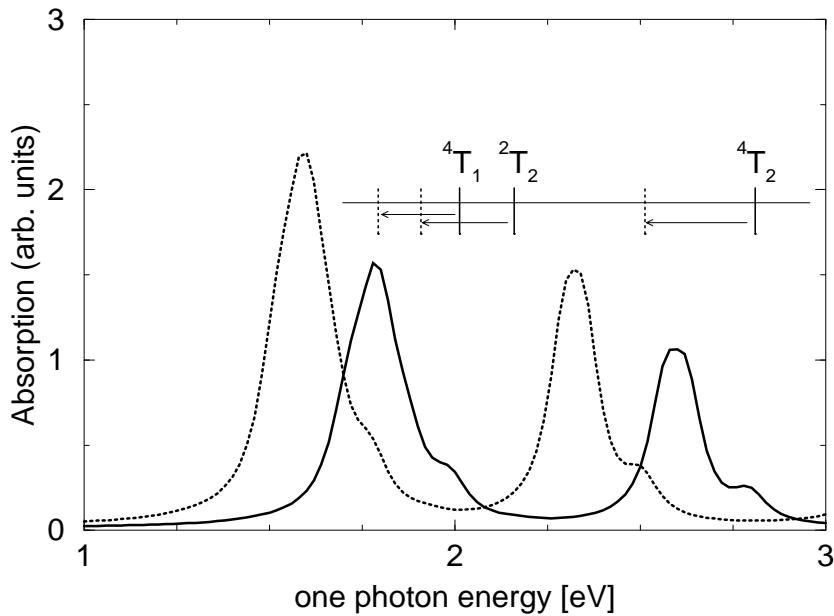}
\caption{
Absorption coefficient as a function of one photon energy.
The solid line represents the calculated spectrum.
Vertical solid lines indicate the energies of the fictitious levels with
disregarding the displacement of Fe atoms. The broken line represents
the calculated spectrum with a different parameter set (see the text).
The broken vertical lines indicate the energies of fictitious levels
corresponding to the latter parameter set.
\label{fig.abs}}
\end{center}
\end{figure}

Finally we comment on the energy levels.
In the present calculation scheme, if we disregard the displacement of Fe atoms
and the spin-orbit interaction, we have the ground state characterized 
as $^{6}A_{1}$, and excited states characterized as $^{4}T_{1}$, $^{2}T_{2}$
and $^{4}T_{2}$ with excitation energies $2.09$, $2.16$ and $2.8$ eV,
respectively.
These states, however, have no $E1$ transition matrix elements from the ground
state. Only after taking account of the displacement of Fe atoms, 
they have finite $E1$ transition matrix elements,
but the energy levels may be shifted and split.
To demonstrate this point, we calculate the absorption 
coefficient $I_{abs}(\omega)$ from the formula
\begin{equation}
 I_{abs}(\omega) \propto \frac{1}{\omega}\sum_{i,{\bf e}}\sum_{n}
   |[T^{E1}({\bf q},{\bf e},i)]_{n,g}|^2 
   \delta(\epsilon_n-\epsilon_g-\hbar\omega).
\end{equation}
Assuming the photon propagates along the $c$ axis, the polarization is 
summed over ${\bf e}$ parallel to the $a$ and $b$ axes.
Figure \ref{fig.abs} shows the calculated spectrum (solid line).
It is clearly seen that the peaks are considerably separate from the
positions of $^{4}T_1$, $^{2}T_2$ and $^{4}T_2$ shown by the vertical
solid lines. Therefore, it does not seem meaningful to assign directly 
these peaks to the fictitious levels which have no $E1$ transition matrix 
elements.
These peak positions depend on parameters such as Slater integrals and 
the hybridization between Fe and O atoms.
The broken line represents the spectrum calculated from a 
different parameter set that $F^2$ and $F^4$ are reduced 
by multiplying a factor $0.81$ instead 
of $0.9$ and $\Delta$ is set to be $4.4$ eV instead of $4.0$ eV. 
Vertical broken lines indicate the corresponding energies of fictitious levels.
In the experiment,\cite{Ogawa2004} a peak is found around $1.6$ eV,
and a shoulder structure around $2.0$ eV. The calculated second peak may
correspond to the shoulder in the experiment.
Further adjustment of parameter sets may improve the calculated spectra,
but we would not seek optimal parameter sets in this paper.

As regards the fiction levels with higher energies, we have
$^{4}A_1$ and $^{4}E$ both with energy $3.68$ eV, $^{2}A_2$ with $3.70$ eV,
$^{2}T_1$ with $3.75$ eV, and so on.
As the same as the low energy levels mentioned above, 
these levels have no $E1$ transition matrix elements from the ground state, 
and would be shifted and split due to the displacement of Fe atoms and the 
spin-orbit interaction. 
Therefore, it would be difficult to assign directly these levels to 
the peaks on the SHG spectra.

\section{Concluding Remarks}

We have analyzed the SHG spectra in a polar ferrimagnet GaFeO$_3$,
using the FeO$_6$ cluster model where the Fe atom is displaced from the center 
of the octahedron. We have fully taken account of the Coulomb interaction 
between the $3d$ states, the spin-orbit interaction on the $3d$ states,
and the hybridization of the oxygen $2p$ states with the $3d$ and $4p$ states.
The $E1$ matrix elements between the $3d$ and $4p$ states are 
evaluated on the Fe atom with the interaction 
$H_{\rm int}=-\frac{1}{c}{\bf j}\cdot{\bf A}$. They are 
larger than those evaluated with the conventional form
$\tilde{H}_{\rm int}=-{\bf P}\cdot{\bf E}$. 
The $E1$ matrix elements between the $3d$ states could become finite through 
the effective hybridization between the $4p$ and $3d$ states 
owing to the breaking
of the space-inversion symmetry.
Note that the same cluster model has been successfully applied to analyzing
not only the optical absorption\cite{Igarashi2009} but also the $K$-edge
x-ray absorption.\cite{Igarashi2010}

In the third-order perturbation with 
$H_{\rm int}=-\frac{1}{c}{\bf j}\cdot{\bf A}$,
we have derived the formula of the probability per unit time
for the process that two photons are absorbed and one photon is emitted.
On the basis of this formula, we have calculated the spectra 
as a function of the two-photon energy in the phase-matching condition.
The calculated SHG intensities exhibit multi-peak structure,
which is in accordance with the experiments.
\cite{Ogawa2004,Eguchi2005,Matsubara2009}
The intensities also signify that another peak structures will be found
outside of the published experimental 
surveys\cite{Ogawa2004,Eguchi2005,Matsubara2009};
one centered around $1.5-2.0$ eV in $I_{aaa}(2 \omega;i)$
and the other centered around $3.0-4.0$ eV in $I_{baa}(2\omega,i)$.
We have found that the amplitude $S_{aaa}(2\omega;i)$ changes its sign 
while $S_{baa}(2\omega;i)$ retains the same value,
when the direction of the local magnetic moment is reversed.
This indicates that $I_{aaa}(2\omega)$ would vanish but
$I_{baa}(2\omega)$ would remain finite in the paramagnetic phase.
Hence our results have reproduced the experimental observations
and $I_{aaa}(2\omega)$ could be called as
the magnetization-induced
SHG intensity, which is completely governed by
the $E$1 process since the $M$1 contribution is negligible.

The quantum-mechanical treatment in the present paper has not directly
been applied to the reflection and the refraction problem,
since the phase coherence is not appropriately treated.
More elaborate treatments using the coherent state might be required.
\cite{Bloembergen}
Researches along this line are relegated to the future study.
On the other hand, the semi-classical treatment is known to cope well with
the reflection and the refraction through the nonlinear susceptibility. 
\cite{Bloembergen,Shen}
We have not directly used the conventional semi-classical method, 
since the form 
$\tilde{H}_{\rm int}=-{\bf P}\cdot{\bf E}$ underestimates considerably
the $E1$ transition. Nonetheless the probability amplitude is very close to 
the conventional form of the nonlinear susceptibility.

\begin{acknowledgments}

This work was partly supported by Grant-in-Aid 
for Scientific Research from the Ministry of Education, Culture, Sport, 
Science, and Technology, Japan.

\end{acknowledgments}

\bibliography{Bibfile}

\end{document}